\documentstyle[12pt]{article}
\def\be {\begin{equation}}
\def\ee {\end{equation}}
\def\ba {\begin{eqnarray}}
\def\ea {\end{eqnarray}}

\def\bi {\begin{itemize}}
\def\ei {\end{itemize}}
\begin{document}
\def\bea{\begin{eqnarray}}
\def\eea{\end{eqnarray}}
\title{\bf  {Interacting new agegraphic Phantom model of dark energy in non-flat universe}}
 \author{M.R. Setare  \footnote{E-mail: rezakord@ipm.ir}
  \\{Department of Science,  Payame Noor University. Bijar. Iran}}
\date{\small{}}

\maketitle
\begin{abstract}
In this paper we consider the new agegraphic model of interacting
dark energy in non-flat universe. We show that the interacting
agegraphic dark energy can be described  by a phantom scalar field.
Then we show this phantomic description of the agegraphic dark
energy and reconstruct the potential of the phantom scalar field.

 \end{abstract}
% \begin{document}

\newpage
% \vspace*{10mm}

\section{Introduction}
According to cosmological observations our universe is undergoing an
accelerating expansion, and the transition to the accelerated phase
has been realized in the recent cosmological past \cite{observ}. In
order to explain this remarkable behavior, and despite the intuition
that this can be achieved only through a fundamental theory of
nature, we can still propose some paradigms for its description.
Thus, we can either consider theories of modified gravity
\cite{ordishov}, or introduce the concept of dark energy which
provides the acceleration mechanism. The dynamical nature of dark
energy, at least in an effective level, can originate from various
fields, such is a canonical scalar field (quintessence)
\cite{quint}, a phantom field, that is a scalar field with a
negative sign of the kinetic term \cite{phant}, or the combination
of quintessence and phantom in a unified model named quintom
\cite{quintom}. The advantage of this combined model is that
although in quintessence the dark energy equation-of-state parameter
remains always greater than $-1$ and in phantom cosmology always
smaller than $-1$, in quintom scenario it can cross $-1$.\\
In addition, many theoretical studies are devoted to understand and
shed light on  dark energy, within the string theory framework. The
Kachru-Kallosh-Linde-Trivedi model \cite{kklt} is a typical example,
which tries to construct metastable de Sitter vacua in the light of
type IIB string theory. Despite the lack of a quantum theory of
gravity, we can still make some attempts to probe the nature of dark
energy according to some principles of quantum gravity. An
interesting attempt in this direction is the so-called ``holographic
dark energy'' proposal
\cite{Cohen:1998zx,Hsu:2004ri,Li:2004rb,holoext}. Such a paradigm
has been constructed in the light of holographic principle of
quantum gravity  \cite{holoprin}, and thus it presents some
interesting features of an underlying theory of dark energy. More
recently a new dark energy model, dubbed agegraphic dark energy has
been proposed \cite{cai1} (see also \cite{zin}), which takes
 into account the Heisenberg uncertainty relation of quantum mechanics together with the gravitational
 effect in general relativity.
\\
In the present paper, we suggest  a correspondence between the new
agegraphic dark energy scenario and the phantom dark energy model.
We show this phantomic description of the interacting new agegraphic
dark energy in non-flat universe, and reconstruct the potential of
the phantom scalar field.

\section{ Interacting agegraphic phantom in non-flat universe }
In this section we obtain the equation of state for the agegraphic
energy density when there is an interaction between agegraphic
energy density $\rho_{\Lambda}$ and a Cold Dark Matter(CDM) with
$w_{m}=0$. The continuity equations for dark energy and CDM are
\begin{eqnarray}
\label{2eq1}&& \dot{\rho}_{\rm \Lambda}+3H(1+w_{\rm \Lambda})\rho_{\rm \Lambda} =-Q, \\
\label{2eq2}&& \dot{\rho}_{\rm m}+3H\rho_{\rm m}=Q.
\end{eqnarray}
The interaction is given by the quantity $Q=\Gamma \rho_{\Lambda}$.
This is a decaying of the agegraphic energy component into CDM with
the decay rate $\Gamma$. Taking a ratio of two energy densities as
$u=\rho_{\rm m}/\rho_{\rm \Lambda}$, the above equations lead to
\begin{equation}
\label{2eq3} \dot{u}=3Hu\Big[w_{\rm \Lambda}+
\frac{1+u}{u}\frac{\Gamma}{3H}\Big]
\end{equation}
Here as in Ref.\cite{WGA}, we choose the following relation for
decay rate
\begin{equation}\label{decayeq}
\Gamma=3b^2(1+u)H
\end{equation}
with  the coupling constant $b^2$.
 Following Ref.\cite{Kim:2005at},
if we define
\begin{eqnarray}\label{eff}
w_\Lambda ^{\rm eff}=w_\Lambda+{{\Gamma}\over {3H}}\;, \qquad w_m
^{\rm eff}=-{1\over u}{{\Gamma}\over {3H}}\;.
\end{eqnarray}
Then, the continuity equations can be written in their standard
form
\begin{equation}
\dot{\rho}_\Lambda + 3H(1+w_\Lambda^{\rm eff})\rho_\Lambda =
0\;,\label{definew1}
\end{equation}
\begin{equation}
\dot{\rho}_m + 3H(1+w_m^{\rm eff})\rho_m = 0\; \label{definew2}
\end{equation}
We consider the non-flat Friedmann-Robertson-Walker universe with
line element
 \be\label{metr}
ds^{2}=-dt^{2}+a^{2}(t)(\frac{dr^2}{1-kr^2}+r^2d\Omega^{2}).
 \ee
where $k$ denotes the curvature of space k=0,1,-1 for flat, closed
and open universe respectively. A closed universe with a small
positive curvature ($\Omega_k\sim 0.01$) is compatible with
observations \cite{ {wmap}, {ws}}. We use the Friedmann equation to
relate the curvature of the universe to the energy density. The
first Friedmann equation is given by
\begin{equation}
\label{2eq7} H^2+\frac{k}{a^2}=\frac{1}{3M^2_p}\Big[
 \rho_{\rm \Lambda}+\rho_{\rm m}\Big].
\end{equation}
Define as usual
\begin{equation} \label{2eq9}\Omega_{\rm
m}=\frac{\rho_{m}}{\rho_{cr}}=\frac{ \rho_{\rm
m}}{3M_p^2H^2},\hspace{1cm} \Omega_{\rm
\Lambda}=\frac{\rho_{\Lambda}}{\rho_{cr}}=\frac{ \rho_{\rm
\Lambda}}{3M^2_pH^2},\hspace{1cm}\Omega_{k}=\frac{k}{a^2H^2}
\end{equation}
According to the new agegraphic dark energy we have following
relation for energy density \cite{cai}
 \be \label{holoda}
  \rho_\Lambda=3n^2M_{p}^{2}\eta^{-2}.
 \ee
 where the numerical factor $3n^2$ is introduced to parameterize some uncertainties, such as the species
of quantum fields in the universe, $\eta$ is conformal time, and
given by \be \label{con}
  \eta=\int\frac{dt}{a}=\int\frac{da}{a^2H}.
 \ee
Using definitions
$\Omega_{\Lambda}=\frac{\rho_{\Lambda}}{\rho_{cr}}$ and
$\rho_{cr}=3M_{p}^{2}H^2$, we get

\begin{equation}\label{hl}
H\eta=\frac{n}{\sqrt{\Omega_{\Lambda}}}
\end{equation}
 Using
Eqs. (\ref{eff}), (\ref{definew1}), (\ref{holoda}), (\ref{hl}), one
can obtain the equation of state as \be \label{stateq}
w_{\Lambda}=-(1-\frac{2}{3na}\sqrt{\Omega_{\Lambda}}+\frac{\Gamma}{3H})\label{eqes}\ee
Then using Eqs.(\ref{decayeq}),(\ref{2eq9}) we can rewrite the above
equation as following \be \label{stateq}
w_{\Lambda}=-(1-\frac{2}{3na}\sqrt{\Omega_{\Lambda}}+\frac{b^2(1+\Omega_{k})}{\Omega_{\Lambda}}),\label{eqes11}\ee
then we can see that $w_{\Lambda}$ can cross the phantom divide if
$\frac{b^2(1+\Omega_{k})}{\Omega_{\Lambda}}>\frac{2}{3na}\sqrt{\Omega_{\Lambda}}$.
This implies that one can generate phantom-like equation of state
from an interacting new agegraphic dark energy model in non-flat
universe only if $\frac{3n b^2}{2}>
\frac{\Omega_{\Lambda}^{3/2}}{(1+\Omega_{k})a}$.\\
 Now we assume that the origin of the dark
energy is a phantom scalar field $\phi$, so \be \label{roph1}
\rho_{\Lambda}=-\frac{1}{2}\dot{\phi}^{2}+V(\phi) \ee \be
\label{roph2} P_{\Lambda}=-\frac{1}{2}\dot{\phi}^{2}-V(\phi) \ee In
this case $w_{\rm \Lambda}$ is given by \be \label{w}w_{\rm
\Lambda}=\frac{-\frac{1}{2}\dot{\phi}^{2}-V(\phi)}{-\frac{1}{2}\dot{\phi}^{2}+V(\phi)}
\ee According to the forms of phantom energy density and pressure
eqs.(\ref{roph1}, \ref{roph2}), one can easily derive the scalar
potential and kinetic energy term as \be \label{v}
V(\phi)=\frac{1}{2}(1-w_{\rm \Lambda})\rho_{\Lambda} \ee \be
\label{phi}\dot{\phi}^{2} =-(1+w_{\rm \Lambda})\rho_{\Lambda} \ee
Differenating Eq.(\ref{2eq7}) with respect to the cosmic time $t$,
one find \be \label{hdot}\dot{H}=\frac{\dot{\rho}}{6H
M_{p}^{2}}+\frac{k}{a^2} \ee where $\rho=\rho_{m}+\rho_{\Lambda}$ is
the total energy density, now using Eqs.(\ref{2eq1}, \ref{2eq2}) \be
\label{doro} \dot{\rho}=-3H(1+w)\rho \ee  where \be
\label{weq}w=\frac{w_{\Lambda}\rho_{\Lambda}}{\rho}=\frac{\Omega_{\Lambda}w_{\Lambda}}{1+\frac{k}{a^2H^2}}
\ee Substitute $\dot{\rho}$ into Eq.(\ref{hdot}), we obtain \be
\label{weq2}
w=\frac{2/3(\frac{k}{a^2}-\dot{H})}{H^2+\frac{k}{a^2}}-1 \ee In a
phantom dominated universe $\dot{H}>0$, from Eq.(\ref{weq2}) one can
see easily that in the $k=0, k=-1$ cases $w<-1$, therefore in this
cases $w_{\Lambda}<-1$ also. For $k=1$, the necessary condition to
obtain $w_{\Lambda}<-1$ is this: $\dot{H}>\frac{1}{a^2}$.\\
 Using
Eqs.(\ref{weq}, \ref{weq2}), one can rewrite the agegraphic energy
equation of state as \be \label{eqes1}
w_{\Lambda}=\frac{-1}{3\Omega_{\Lambda}H^{2}}(2\dot{H}+3H^2+\frac{k}{a^2})
\ee Substitute the above $w_{\Lambda}$ into Eqs.(\ref{v},
\ref{phi}), we obtain \be \label{v1} V(\phi)=\frac{M_{p}^{2}}{2}
[2\dot{H}+3H^2(1+\Omega_{\Lambda})+\frac{k}{a^2}]\ee  \be
\label{phi2}\dot{\phi}^{2}
=M_{p}^{2}[2\dot{H}+3H^2(1-\Omega_{\Lambda})+\frac{k}{a^2}]\ee In
the spatially flat case, $k=0$,and  $\Omega_{\Lambda}=1$, in this
case the Eqs.(\ref{v1}, \ref{phi2}) are exactly Eq.(6) in \cite
{odi1} if we consider $\omega(\phi)=-1$. In similar to the \cite
{{odi1}, {odi2}}, we can define $\dot{\phi}^{2}$ and $V(\phi)$ in
terms of single function $f(\phi)$ as \be \label{v2}
V(\phi)=\frac{M_{p}^{2}}{2}
[2f'(\phi)+3f^{2}(\phi)(1+\Omega_{\Lambda})+\frac{k}{a^2}]\ee \be
\label{phi3}1=M_{p}^{2}
[2f'(\phi)+3f^{2}(\phi)(1-\Omega_{\Lambda})+\frac{k}{a^2}]\ee In the
spatially flat case the Eqs.(\ref{v2}, \ref{phi3}) solved only in
case of presence of two scalar potentials  $V(\phi)$, and
$\omega(\phi)$. Here we have claimed that in the presence of
curvature term $\frac{k}{a^2}$, Eqs.(\ref{v2}, \ref{phi3}) may be
solved with  potential  $V(\phi)$ ( To see the general procedure for
such type calculations refer to \cite{{odi1},{odi2}}). Hence, the
following solution are obtained \be \label{sol} \phi=t, \hspace{1cm}
H=f(t) \ee One can check that the solution (\ref{sol}) satisfies the
following scalar field equation \be
\label{phieq}-\ddot{\phi}-3H\dot{\phi}+V'(\phi)=0 \ee Therefore by
the above condition, $f(\phi)$ in our model must satisfy following
relation \be \label{coneq} 3f(\phi)=V'(\phi)\ee
 In the other hand, using
Eqs.(\ref{holoda}),(\ref{eqes11}), (\ref{v}), and (\ref{phi}) we
have \be \label{v4}V(\phi)=(3-\frac{\sqrt{\Omega_{\rm
\Lambda}}}{na}+\frac{3b^2(1+\Omega_k)}{2\Omega_{\rm
\Lambda}})M^{2}_{p}H^2\Omega_{\rm \Lambda} \ee \be \label{phi1}
\dot{\phi}=(3b^2(1+\Omega_k)-\frac{2}{na}\Omega_{\rm
\Lambda}^{3/2})^{1/2}M_pH \ee Using Eq.(\ref{phi1}), we can rewrite
Eq.(\ref{v4}) as \be \label{v5}V(\phi)=3 M_{p}^{2} H^2\Omega_{\rm
\Lambda}+\frac{\dot{\phi}^{2}}{2},\ee or in another form as
following \be \label{v5}V(\phi)=3 M_{p}^{2} f^{2}(\phi)\Omega_{\rm
\Lambda}+\frac{1}{2}\ee Then, from Eqs.(\ref{v2}, \ref{v5}), we get
\be \label{keq}\frac{k}{a^{2}}=3f^{2}(\phi)(\Omega_{\rm
\Lambda}-1)-2f'(\phi)+\frac{1}{M_{p}^{2}} \ee In the other hand,
using Eqs.(\ref{eqes11}),(\ref{eqes1}), and (\ref{sol}), one can
obtain \be \label{second} \frac{k}{a^2}=3\Omega_{\rm
\Lambda}f^{2}(\phi)(1-\frac{2\sqrt{\Omega_{\rm
\Lambda}}}{3na}+\frac{b^2(1+\Omega_k)}{\Omega_{\rm
\Lambda}})-2f'(\phi)-3f^{2}(\phi)  \ee Now, using
Eqs.(\ref{keq}),(\ref{second}) we obtain \be \label{omega}
\Omega_{\rm
\Lambda}=(\frac{na}{2})^{2/3}(3b^2(1+\Omega_k)-\frac{1}{M^{2}_{p}f^{2}(\phi)})^{2/3},
\ee where \be \label{aeq} a=e^{\int f(\phi)d\phi} \ee
 Substitute the above $\Omega_{\rm \Lambda}$ into
Eq.(\ref{v5}), we obtain the scalar potential as following\be
\label{v6}V(\phi)=3 M_{p}^{2}f^{2}(\phi)(\frac{n}{2})^{2/3}
e^{\frac{2}{3}\int
f(\phi)d\phi}(3b^2(1+\Omega_k)-\frac{1}{M^{2}_{p}f^{2}(\phi)})^{2/3}+\frac{1}{2}.
\ee
 \section{Conclusions}
Although a complete theory of quantum gravity in not established
yet, one can make some attempts to investigate the nature of dark
energy according to some principles of quantum gravity. The
agegraphic and new agegraphic models are such examples. In this
paper we have associated the interacting new agegraphic dark energy
in non-flat universe with a phantom scalar field. We have shown that
the new agegraphic dark energy can be described
 by the phantom in a certain way. Then a
correspondence between the new agegraphic dark energy and phantom
has been established, and the potential of the agegraphic phantom
has been reconstructed.


\begin{thebibliography}{99}
\bibitem{observ}
A. G.  Riess {\it{et al.}} [Supernova Search Team Collaboration],
Astrophys. J. {\bf 607}, 665 (2004);
 S.
Perlmutter {\it{et al.}} [Supernova Cosmology Project
Collaboration], Astrophys. J. {\bf 517}, 565 (1999);
 D. N. Spergel {\it{et al.}}, Astrophys.
J. Suppl. {\bf 148}, 175 (2003); S. W. Allen, {\it{et al.}}, Mon.
Not. Roy. Astron. Soc. {\bf 353}, 457 (2004).


\bibitem{ordishov}
S.Nojiri and S.~D.~Odintsov, Phys. Rev. D {\bf{68}}, 123512 (2003);
 S.Nojiri and S.~D.~Odintsov, Int. J. Geom. Meth. Mod. Phys.
{\bf{4}}, 115 (2007);
%\cite{Setare:2008hm}
  M.~R.~Setare and E.~N.~Saridakis,
  %``Correspondence between Holographic and Gauss-Bonnet dark energy models,''
  Phys.\ Lett.\  B {\bf 670}, 1 (2008)
  [arXiv:0810.3296 [hep-th]].
  %%CITATION = PHLTA,B670,1;%%



\bibitem{quint}
B.~Ratra and P.~J.~E.~Peebles, Phys.\ Rev.\ D {\bf 37}, 3406 (1988);
C.~Wetterich, Nucl.\ Phys.\ B {\bf 302}, 668 (1988); A.~R.~Liddle
and R.~J.~Scherrer, Phys.\ Rev.\ D {\bf 59}, 023509 (1999)
[arXiv:astro-ph/9809272]; I.~Zlatev, L.~M.~Wang and
P.~J.~Steinhardt, Phys.\ Rev.\ Lett.\ {\bf 82}, 896 (1999);
Z.~K.~Guo, N.~Ohta and Y.~Z.~Zhang, Mod.\ Phys.\ Lett.\  A {\bf 22},
883 (2007).

\bibitem{phant} R. R. Caldwell, Phys.
Lett. B {\bf{545}}, 23 (2002); R.~R.~Caldwell, M.~Kamionkowski and
N.~N.~Weinberg, Phys. Rev. Lett. {\bf 91}, 071301 (2003); S. Nojiri
and S. D. Odintsov, Phys. Lett. B {\bf 562}, 147 (2003)
[arXiv:hep-th/0303117]; V. K. Onemli and R. P. Woodard, Phys.\ Rev.\
D {\bf 70}, 107301 (2004) [arXiv:gr-qc/0406098]; M. R. Setare, Eur.
Phys. J. C {\bf 50}, 991 (2007).


\bibitem{quintom}
B.~Feng, X.~L.~Wang and X.~M.~Zhang, Phys.\ Lett.\  B {\bf 607}, 35
(2005);
  %%CITATION = PHLTA,B607,35;%%
Z. K. Guo, {\it{et al.}}, Phys. Lett. B {\bf 608}, 177 (2005); M.-Z
Li, B. Feng, X.-M Zhang, JCAP, 0512, 002 (2005); B. Feng, M. Li,
Y.-S. Piao and X. Zhang, Phys. Lett. B {\bf 634}, 101 (2006); M. R.
Setare, Phys. Lett. B {\bf 641}, 130 (2006); W. Zhao and Y. Zhang,
Phys. Rev. D {\bf73}, 123509 (2006);
 M. R.
Setare, J. Sadeghi, and A. R. Amani, Phys. Lett. B {\bf 660}, 299
(2008); J. Sadeghi, M. R. Setare, A. Banijamali and F. Milani, Phys.
Lett. B {\bf 662}, 92 (2008);
%\cite{Setare:2008pz}
  M.~R.~Setare and E.~N.~Saridakis,
  Phys.\ Lett.\  B {\bf 668}, 177 (2008);
  %%CITATION = PHLTA,B668,177;%%
%\cite{Setare:2008dw}
  M.~R.~Setare and E.~N.~Saridakis,
  [arXiv:0807.3807 [hep-th]];
  %%CITATION = ARXIV:0807.3807;%%
 %\cite{Setare:2008si}
  M.~R.~Setare and E.~N.~Saridakis,
  JCAP {\bf 0809}, 026 (2008).
\bibitem{kklt}
  S.~Kachru, R.~Kallosh, A.~Linde and S.~P.~Trivedi,
  Phys.\ Rev.\ D {\bf 68}, 046005 (2003).

\bibitem{Cohen:1998zx}
  A.~G.~Cohen, D.~B.~Kaplan and A.~E.~Nelson,
  Phys.\ Rev.\ Lett.\  {\bf 82}, 4971 (1999);
  P.~Horava and D.~Minic,
  Phys.\ Rev.\ Lett.\  {\bf 85}, 1610 (2000);
   S.~D.~Thomas,
  Phys.\ Rev.\ Lett.\  {\bf 89}, 081301 (2002).

\bibitem{Hsu:2004ri}
  S.~D.~H.~Hsu,
  Phys.\ Lett.\ B {\bf 594}, 13 (2004).

\bibitem{Li:2004rb}
  M.~Li,
  Phys.\ Lett.\ B {\bf 603}, 1 (2004);
  D.~Pavon and W.~Zimdahl,
  Phys.\ Lett.\ B {\bf 628}, 206 (2005).

\bibitem{holoext}
  K.~Enqvist and M.~S.~Sloth,
  Phys.\ Rev.\ Lett.\  {\bf 93}, 221302 (2004);
  K.~Ke and M.~Li,
  Phys.\ Lett.\ B {\bf 606}, 173 (2005);
  Q.~G.~Huang and M.~Li,
  JCAP {\bf 0503}, 001 (2005);
    E.~Elizalde, S.~Nojiri, S.~D.~Odintsov and P.~Wang,
  Phys.\ Rev.\ D {\bf 71}, 103504 (2005);
  B.~Wang, Y.~Gong and E.~Abdalla,
  Phys.\ Lett.\ B {\bf 624}, 141 (2005);
    S.~Nojiri and S.~D.~Odintsov,
  Gen.\ Rel.\ Grav. {\bf 38}, 1285 (2006);
  H.~Kim, H.~W.~Lee and Y.~S.~Myung,
  Phys.\ Lett.\ B {\bf 632}, 605 (2006);
  B.~Hu and Y.~Ling,
  Phys.\ Rev.\ D {\bf 73}, 123510 (2006);
  H.~Li, Z.~K.~Guo and Y.~Z.~Zhang,
  Int.\ J.\ Mod.\ Phys.\ D {\bf 15}, 869 (2006);
   M.~R.~Setare,
  Phys.\ Lett.\ B {\bf 642}, 1 (2006);
  M.~R.~Setare, Phys. Lett. B {\bf 642}, 421 (2006);
  M. R. Setare, Phys. Lett. B {\bf 644}, 99 (2007);
M. R. Setare, J. Zhang and X. Zhang, JCAP {\bf 0703} 007 (2007);
 M. R. Setare, Phys. Lett.  B {\bf 648}, 329 (2007); M. R.
Setare, Phys. Lett. B {\bf 654}, 1 (2007); W. Zhao, Phys. Lett. B
{\bf 655}, 97, (2007); M. Li, C. Lin and Y. Wang, JCAP {\bf 0805},
023 (2008).

\bibitem{holoprin}
G.~'t Hooft,
  [arXiv:gr-qc/9310026];
 L.~Susskind,
  J.\ Math.\ Phys.\  {\bf 36}, 6377 (1995).
\bibitem{cai1}R. G. Cai, Phys. Lett. B {\bf 657}, 228, (2007).
\bibitem{zin}I. P. Neupane, Phys. Lett. B {\bf 673}, 111, (2009);
K. Y. Kim, H. W. Lee, Y. S. Myung, Phys. Lett. B {\bf 660}, 118,
(2008; J. Zhang, X. Zhang, H. Liu, Eur. Phys. J. {\bf C54}, 303,
(2008).
\bibitem{WGA} B. Wang, Y. Gong, and E. Abdalla, Phys. Lett. B {\bf 624}, 141
(2005).
\bibitem{Kim:2005at} H. Kim, H. W. Lee and Y. S. Myung,
Phys. Lett. B {\bf 632}, 605, (2006).
\bibitem{wmap}C. L. Bennett
et al., Astrophys. J. Suppl. 148, 1 (2003).
\bibitem{ws}M. Tegmark et al., Phys.\ Rev.\ D {\bf 69}, 103501 (2004).
\bibitem{cai}H. Wei, R. G. Cai, Phys. Lett. B {\bf660}, 113, (2008);
K. Y. Kim, H. W. Lee, Y. S. Myung, M. I. Park, Mod. Phys. Lett. {\bf
A23}, 3049, (2008);  J .P Wu, D. Z. Ma, Y. Ling, Phys. Lett. {\bf
B663}, 152, (2008); J. Cui, L. Zhang, J. Zhang, X. Zhang,
arXiv:0902.0716 [astro-ph].
\bibitem{odi1}S. Nojiri,  and S. D. Odintsov, Gen. Rel. Grav. {\bf 38}, 1285,
(2006).
\bibitem{odi2}S. Capozziello, S. Nojiri,  and S. D. Odintsov,  Phys. Lett. {\bf B632}, 597, (2006);
S. Nojiri,  and S. D. Odintsov, hep-th/0611071; S. Nojiri, S. D.
Odintsov, and H. Stefancic, Phys. Rev. {\bf D74}, 086009, (2006).


\end{thebibliography}
\end{document}